\documentclass{article}
\usepackage{graphicx} 
\usepackage{xcolor}
\usepackage{setspace}
\usepackage{multirow}
\usepackage[normalem]{ulem}
\useunder{\uline}{\ul}{}
\usepackage{anyfontsize}
\usepackage{longtable}
\usepackage{array}
\usepackage{multirow}
\usepackage{hyperref}
\usepackage[utf8]{inputenc}
\usepackage{booktabs}
\usepackage{authblk}
\begin{document}

\title{Programmable Virtual Humans Toward Human Physiologically-Based Drug Discovery}
\author{%
  You Wu$^{1,*}$, Philip E. Bourne$^{2}$, Lei Xie$^{1,3,4,5,*}$
  \date{} 
  \\
  \small
  $^{1}$Ph.D. Program in Computer Science, The Graduate Center, The City University of New York, New York, New York, USA \\
  $^{2}$School of Data Science \& Department of Biomedical Engineering, University of Virginia, Charlottesville, Virginia, USA \\
  $^{3}$School of Pharmacy and Pharmaceutical Sciences \& Center for Drug Discovery, Northeastern University, Boston, MA, USA \\
  $^{4}$Department of Computer Science, Hunter College, The City University of New York, New York, New York, USA \\
  $^{5}$Helen \& Robert Appel Alzheimer’s Disease Research Institute, Feil Family Brain \& Mind Research Institute, Weill Cornell Medicine, Cornell University, New York, New York, USA \\
  $^{*}$Correponding authors: ywu1@gradcenter.cuny.edu, le.xie@northeastern.edu  \\
}
\maketitle

\begin{abstract}
 Artificial intelligence (AI) has sparked immense interest in drug discovery, but most current approaches only digitize existing high-throughput experiments. They remain constrained by conventional pipelines. As a result, they do not address the fundamental challenges of predicting drug effects in humans. Similarly, biomedical digital twins, largely grounded in real-world data and mechanistic models, are tailored for late-phase drug development and lack the resolution to model molecular interactions or their systemic consequences, limiting their impact in early-stage discovery. This disconnect between early discovery and late development is one of the main drivers of high failure rates in drug discovery. The true promise of AI lies not in augmenting current experiments but in enabling virtual experiments that are impossible in the real world: testing novel compounds directly in silico in the human body. Recent advances in AI, high-throughput perturbation assays, and single-cell and spatial omics across species now make it possible to construct programmable virtual humans: dynamic, multiscale models that simulate drug actions from molecular to phenotypic levels. By bridging the translational gap, programmable virtual humans offer a transformative path to optimize therapeutic efficacy and safety earlier than ever before. This perspective introduces the concept of programmable virtual humans, explores their roles in a new paradigm of drug discovery centered on human physiology, and outlines key opportunities, challenges, and roadmaps for their realization.
\end{abstract}


\section{Introduction}

The recent success of artificial intelligence (AI) in image and video generation, natural language processing (NLP), and structural biology has sparked enthusiasm for AI's application to drug discovery and development. Most efforts in AI-based drug discovery currently focus on the conventional one-drug-one-target paradigm, where a single disease associated gene is identified and prioritized as drug target and small molecules are screened or designed to modulate the target. However, the success rate of target-based drug discovery is extremely low for complex diseases due to lack of efficacy or unforeseen side effects in clinic\cite{sams2005target}. It is not surprising since target-based drug discovery oversimplifies the complex, multi-genic nature of diseases, especially chronic and multi-factorial conditions like cancer, neurodegenerative disorders, and autoimmune diseases. In many cases, the understanding of target's role in disease biology is incomplete. The initial hypothesis about a target’s involvement in a disease seldom hold up under rigorous clinical testing. 

The high failure rate and costs of target-based drug discovery have renewed interest in phenotype-based drug discovery using cell line or organoid models. However, phenotype-based drug discovery has several inherent drawbacks. Firstly, disease models rarely accurately replicate human disease conditions. Secondly, the pharmacokinetics and pharmacodynamics \textit{in vitro} of a compound are significantly different from its \textit{in vivo} drug actions in the human body. Finally, molecular targets and mechanisms of action are often unknown, leading to regulatory challenges. Existing work in applying AI to phenotype-based screening mainly focus on solving problems of high-throughput phenotype screening using the disease model, but pay little attentions to addressing the fundamental challenges in the discrepancy between disease models and humans.   

As highlighted by Bender et al. \cite{bender2021artificial}, a significant challenge in target-based or phenotype-based drug discovery is the translational gap between molecules optimized for binding affinities to a single target or potency in a disease model and therapeutics that demonstrate clinical efficacy and safety in humans. In a retrospective, the technical advances from combinatorial library and DNA-encoded library (DEL) to high-content screening have not significantly increased the productivity and reduced the cost of drug discovery and development for complex diseases. Without a paradigm shift in drug discovery and development that addresses the translational gaps, the success of AI in each isolated steps in current drug discovery process may lead to faster and cheaper failures rather than increased success rates in clinic. For instance, a lead compound effectively optimized by state-of-the-art structure-based reinforcement learning or diffusion model might target a gene that is neither a causal nor sufficient driver of a disease. Even if the primary target is correctly prioritized, an unexpected off-target binding of the optimized lead may occur, leading to side effects and possible withdrawal in clinic. Similarly, a drug candidate optimized by a powerful machine learning model using iPSC screening may not be effective or safe in humans, and an optimal clinical trial design might still fail if the underlying assumptions of target's roles in disease biology are incorrect. 

Ideally, drugs should be screened or designed by directly optimizing their clinical effects in the human body even we know little about complex disease mechanisms. However, experimentation on humans is infeasible and unethical. Human digital twins that are virtual replicas of human beings have been proposed to simulate drug actions in the human body for healthcare, clinical trials and personalized medicine\cite{bjornsson2020digital,laubenbacher2024digital,armeni2022digital,tang2024roadmap}. These digital twin technologies are typically based on human clinical data or mechanism-based modeling at the tissue or organ level. The success of these approaches relies on sufficient human data or detailed knowledge of disease mechanisms and drug actions. However, for new extraneous molecules, neither human data nor pharmacological information such as pharmacokinetics is available. Thus, existing digital twins cannot reliably predict their clinical outcomes in the human body at the early stage of drug discovery. A new type of programmable virtual humans (PVH) is needed to predict the physiological effects of new unseen molecules and enable the inverse design of novel therapeutics to revert disease phenotypes to a healthy state even lacking of comprehensive understanding of disease etiology.

In this perspective, we first propose the PVH that can simulate the entire pharmaceutical process of a new molecule within the human body, aiming to a new drug discovery paradigm that is based on programmable human physiology. We then outline the opportunities, challenges, and road maps involved in developing PVH by integrating artificial intelligence, knowledge engineering, systems biology, and biophysics.

\begin{figure}
    \vspace{-3.5cm}
    \hspace{-3.5cm}
\includegraphics[width=1.4\linewidth]{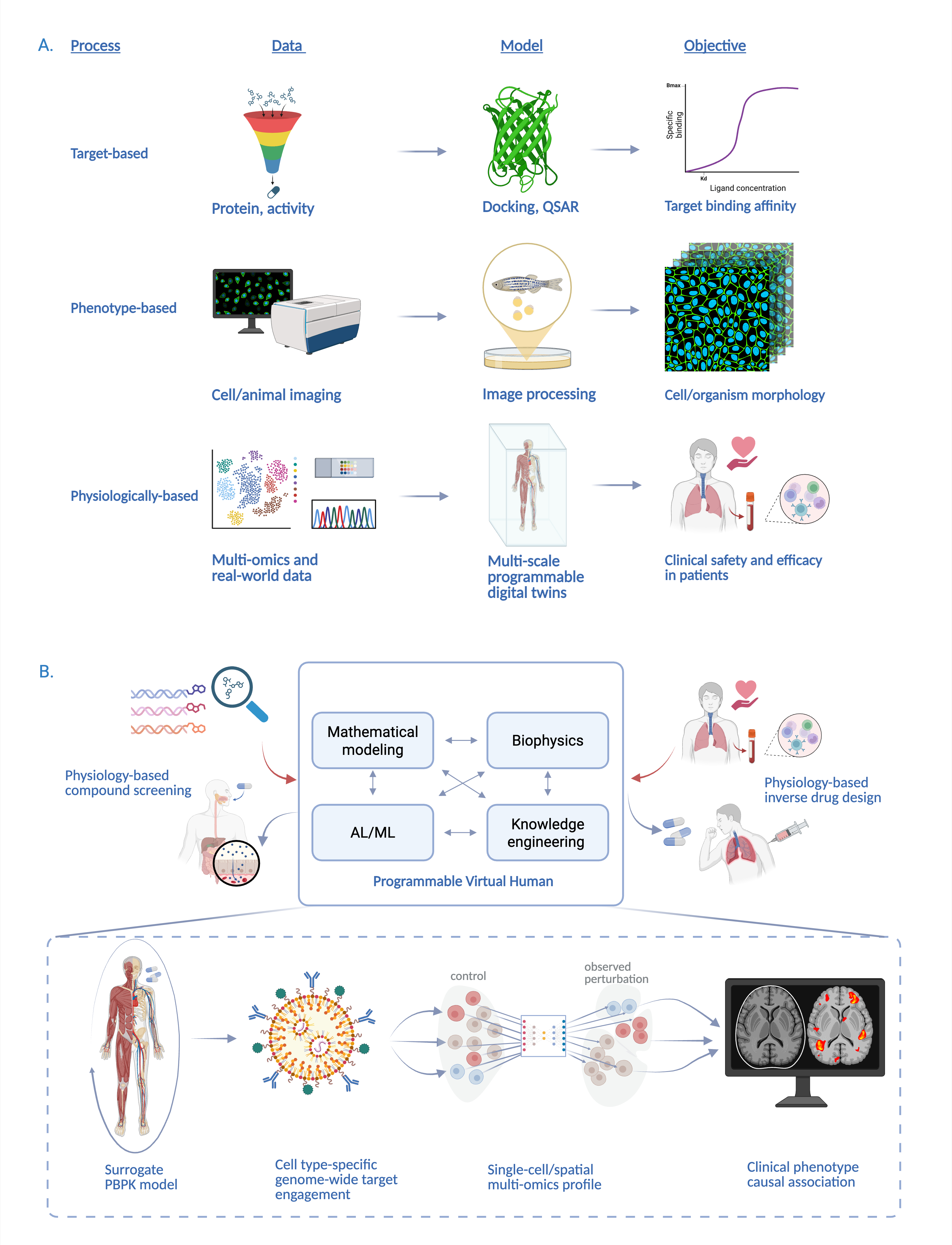}
    \caption{\small (A) Overview of three drug discovery paradigms: target-based, phenotype-based, and physiology-based. Each leverages distinct information, such as protein activity, imaging, or multi-omics, to build computational models (e.g., QSAR, image processing, digital twins) with objectives ranging from binding affinity prediction to morphological changes and clinical safety assessment. (B) The proposed "PVH (Programmable Virtual Human)" integrates mathematical modeling, biophysics, machine learning, and knowledge engineering to enable physiology-based compound screening and inverse drug design. The workflow involves simulating pharmacokinetics, predicting cell-type-specific target engagement, analyzing single-cell/spatial omics under perturbation, and linking molecular changes to clinical phenotypes.}
    \label{fig:main}
\end{figure}

\section{Programmable virtual humans for human physiologically based drug discovery}

A computational platform for drug discovery comprises three key components: process, data, and method. For example, in a conventional target-based drug discovery process, the necessary data includes Genome-wide Association Studies (GWAS) and other omics data, the protein structure of the target or the binding affinities of a compound library. The corresponding methods typically involve bioinformatics, systems biology and machine learning approach for target identification and protein-ligand docking or quantitative structure-activity relationship (QSAR) modeling for hit discovery and lead optimization. 

The goal of the PVH is to enable a new paradigm of human physiologically based drug discovery, overcoming the limitations of the existing drug discovery and development process, as illustrated in Figure \ref{fig:main}. Physiologically based drug discovery directly optimizes a drug candidate to maximize therapeutic efficacy and minimize side effects in \textbf{patients}. It will bridge the translational gaps from target identification and validation, lead discovery and optimization, pre-clinical animal studies, clinical trials, to clinical use. Completely different from the conventional linear drug discovery process where success in each individual step may not translate to final clinical success. 
This approach allows us to design drugs analogous to using ChatGPT to generate an image from a text prompt. Given a patient's disease state and healthy state represented by a patient's PVH and a healthy reference one, respectively, they can be used as "prompts" to generate a molecule that reverts the patient's disease state to a healthy state.

To enable physiologically based drug discovery using the PVH, the required data must not only characterize human physiology but also provide a mechanistic understanding of drug actions in the human body. The availability of various omics data provides a solid foundation for developing the PVH. For example, chemoproteomics data enables the characterization and prediction of a chemical’s target engagement in a cellular context\cite{mitchell2023proteome, reinecke2024chemical, offensperger2024large}. Chemical-perturbed omics data supports predictive modeling of chemically induced molecular phenotype changes\cite{wu2024ai}. Multi-modal biobank data facilitates linking molecular characterizations to clinical observations. To effectively utilize heterogeneous multi-omics and real-world data for the PVH, it is crucial to integrate them across multiple dimensions: molecular modalities (DNA, RNA, proteins, and metabolites), biological hierarchies (molecules, cells, tissues, organs, humans, and populations), and species (from model organisms to humans)\cite{wu2024ai}. 

Current mainstream computational tools for drug discovery have significant limitations. Most datasets are used separately and for individual steps in the drug discovery process. Omics data is primarily applied to target identification and prioritization. Structure-based drug design approaches, such as AlphaFold 3, do not account for target engagement within a cellular context. Optimizing binding affinity alone does not guarantee therapeutic efficacy in patients. PBPK models and medical digital twins rely on preclinical or clinical data, which is often unavailable in the early stages of drug discovery. The PVH unifies data and process silos in the existing AI-powered drug discovery pipeline to end-to-end optimize drug effects. It simulates physiologically based pharmacokinetics (PBPK) and pharmacodynamics for \textit{novel compounds}, maps molecular phenotype responses at the single-cell level, models cell-cell interactions across tissues and organs, and connects drug-induced cellular phenotypes to organism-level diagnoses and therapeutic biomarkers. Additionally, the PVH integrates data-driven modeling with mechanism-based approaches, such as mathematical modeling, biophysics, and knowledge engineering, to address data scarcity and domain shift challenges.

\section{Opportunities on developing PVH}

The success of PVH hinges on two critical factors: the availability of data that accurately captures the molecular components and their interactions within human physiological systems, and advanced modeling techniques capable of integrating heterogeneous, noisy, sparse, and high-dimensional data to make reliable and interpretable predictions on unseen cases. Recent advances in omics technology, AI, systems biology, and biophysics have paved the way for developing the PVH to simulate drug actions in the human body. Unlike classic machine learning, deep learning offers the unique capability of using both labeled and unlabeled multi-modal data, end-to-end optimizing clinical outcomes across biological levels from DNAs to RNAs to proteins to metabolites and organismal hierarchies from molecules to cells to tissues to organs to the human body,  and incorporating a prior scientific knowledge — precisely what are needed to construct the PVH that can simulate the entire life cycle of drug actions and predict human responses to novel chemical perturbations.

\subsection{Simulating the fate of unseen molecules in the human body}

One of the central tasks of the PVH is to predict the pharmacokinetics and pharmacodynamics of new molecules. Physiologically Based Pharmacokinetics (PBPK) modeling and Quantitative Systems Pharmacology (QSP) have emerged as powerful tools in drug development and regulatory decision-making\cite{lin2022applications}. However, because both PBPK and QSP rely on solving differential equations derived from mechanistic models, they have limited capabilities in predicting physiological drug concentrations and effects for new molecules. Recently, several studies have applied machine learning to enhance the predictive power of PBPK and QSP\cite{chou2023machine}. Tremendous efforts have been devoted to developing chemical foundation models to predict molecular properties relevant to pharmacokinetics\cite{mendez2024mole}. A particularly promising approach is Physics-Informed Neural Networks (PINNs)\cite{cuomo2022scientific}. It has shown promising results in climate and weather prediction\cite{verma2024climode}. By incorporating physical laws into neural networks, expressed through differential equations or other mathematical formulations, PINNs offer an attractive solution to improve data efficiency, generalization, and interpretability.

After determining the concentration of a molecule in a cell, the next crucial question is how this molecule interacts with various cellular components (DNAs, RNAs, proteins, etc.). Since most drugs are designed to target proteins, predicting the binding poses, thermodynamics, kinetics, and functional selectivity of genome-wide protein-chemical interactions is essential\cite{xie2012novel}. Recent advancements in protein structure prediction have sparked significant progress in predicting protein-small molecule complexes, particularly using large language model\cite{esm2_lin2023evolutionary} and diffusion models\cite{alphafold3_abramson2024accurate}. Other advanced deep learning techniques, including semi-supervised learning\cite{mmaple}, meta-learning\cite{mmaple, portalcg_cai2023end}, transfer learning\cite{wang2024leveraging}, and contrastive learning\cite{singh2023contrastive}, have been explored to enhance genome-wide protein-chemical interaction predictions. Ligand binding kinetics are more closely correlated with drug efficacy and toxicity in the human body than binding affinities\cite{chiu2016toward}. However, predicting binding kinetics using machine learning remains challenging due to the scarcity of kinetic data. Molecular Dynamics (MD) simulations of binding kinetics are constrained by the long time scales required, ranging from minutes to hours. The development of reliable machine-learned force fields\cite{frank2024euclidean} and conformational samplings\cite{janson2023direct} may enable more efficient and accurate predictions of ligand binding kinetics. 

Predicting functional selectivity of ligand binding (agonist, partial agonist, and antagonist, etc.) and biased signaling remains a missing piece in simulating drug actions\cite{cai2022deepreal}. It is particularly important for G-protein coupled receptors (GPCRs), one of the most studied drug targets. Because the functional selectivity is strongly associated with protein conformational dynamics, it is interesting to integrate machine learning with MD simulations. Additionally, chemical-induced omics profiles are comprehensive readouts of ligand functional selectivity and biased signaling. Thus, integration of physics-based simulations with machine learning models that predict ligand-induced omics profiles could be a promising direction. 

\subsection{Modeling multi-facets of cell states}

The PVH must accurately capture cell state changes in response to molecular perturbations across diverse genetic and environmental backgrounds. A cell state can be characterized by its molecular landscape, which includes DNA, RNA, proteins, metabolites, and other biomolecules. Recent advances in single-cell omics techniques enable the detailed characterization of cell states, including genomics\cite{cuomo2023single}, epigenomics\cite{kelsey2017single}, transcriptomics\cite{jovic2022single}, proteomics\cite{bennett2023single} (including protein post-translational modifications\cite{orsburn2022insights}), metabolomics\cite{vicari2024spatial}, lipidomics\cite{li2021single}, glycomics\cite{marie2024native}, microbiomics\cite{jia2024single}, morphology\cite{zhao2023single}, and others. Leveraging these data, systems biology and machine learning have made significant strides in identifying molecular drivers of phenotypes\cite{kunes2024supervised}, reconstructing gene regulatory networks\cite{chang2024single}, mapping directional protein-protein interaction networks\cite{pirak2024d}, building genome-scale metabolic networks\cite{kundu2024machine}, and uncovering temporal dynamics of biological processes\cite{carilli2024biophysical}. These advancements provide a solid foundation for developing the PVH. 

Each individual omics data provides only partial insights into cellular functions. To accurately simulate how a cell responses to external perturbations, it is necessary to integrate various types of omics data across biological levels from DNAs to RNAs to proteins to metabolites to cellular phenotypes. Multi-modal machine learning is a powerful tool for achieving this integration\cite{steyaert2023multimodal}. In particular, deep learning enables the use of unlabeled data to develop foundation models for DNAs\cite{nguyen2024sequence}, RNAs\cite{bian2024development}, and proteins\cite{esm2_lin2023evolutionary}. It also allows for the fusion of diverse data modalities into a unified representation space\cite{saturn_rosen2024toward} and the simulation of biological information flows across different biological levels\cite{he2022cross, transpro_wu2023hierarchical}. A biology-inspired, end-to-end deep learning framework holds significant potential for predictive modeling of genotype-environment-phenotype relationships\cite{wu2024ai}.

The human body is composed of diverse cell types organized hierarchically into tissues, organs, and systems, with cells communicating through chemical signals. Dysfunctions in one part of the body can influence or be influenced by other parts. For example, the gut microbiome impacts the central nervous system\cite{sharon2016central}. Hypertension or heart disease can affect kidney function, while chronic kidney disease can lead to cardiovascular complications, creating a vicious cycle of organ dysfunction\cite{said2014link}. Advances in spatial omics techniques now allow characterization and organization of organome-, cellulome, and genome-wide cellular processes\cite{bressan2023dawn}. For instance, Lilja et al. developed a multi-organ, multicellular disease model and identified pro- and anti-inflammatory upstream regulators for personalized medicine\cite{lilja2023multi}. Wu et al. used deep learning to elucidate the molecular basis of microbiome-human interactions associated with various diseases\cite{mmaple}. These examples underscore the interconnected nature of human biology, where multi-scale approaches are vital for understanding and treating complex diseases.

\subsection{Predicting clinical responses to novel perturbations}

One of the most exciting advancements enabling the PVH is the development of perturbation functional genomics and image-based profiling techniques\cite{abudayyeh2024programmable}, notably perturb-seq\cite{xu2024dissecting,yao2024scalable}, epigenome editing\cite{mccutcheon2024epigenome}, drug-seq\cite{ye2018drug}, and cell paintings\cite{chandrasekaran2021image} as well as human microphysiolocal systems (MPS) for genetic and drug responses\cite{southard2024comprehensive}. These techniques systematically alter specific genes or molecular pathways within cells to observe the resulting phenotypic changes. With sufficient perturbation data, it becomes possible to reverse-engineer the complexities of the human physiology. 

Recent advances in AI have produced several innovative approaches for predicting and modeling cellular responses to perturbations. For example, ChemCPA uses an encoder-decoder architecture with adversarial training to transfer bulk RNA-seq data to single-cell contexts, disentangling molecular attributes for detailed gene expression analysis\cite{chemcpa_hetzel2022predicting}. GEARS employs a graph neural network (GNN) on gene interaction networks to predict transcriptional responses to multi-gene perturbations, enabling insights into unseen gene combinations\cite{gears_roohani2023predicting}. MultiDCP incorporates multi-omics data to predict dosage-specific drug responses including cell viability and transcriptomics\cite{multidcp_wu2022deep}, generalizing across novel drugs and cell lines. More recent research has expanded into digital twin development by combining cell behavior ontologies with single-cell and spatial omics data through Agent-based models (ABMs)\cite{johnson2023digitize}.

While perturbation functional genomics and image data and MPS are invaluable for developing the PVH, they are not obtained directly from patients. Even the most advanced models, such as organ-on-a-chip, cannot fully replicate human physiology\cite{pang2024tackling}. Therefore, translating the knowledge gained from disease models to human systems is crucial. The integration of AI, systems biology, and biophysics again offers significant potential to achieve this. First, accurate PBPK modeling for novel chemicals, as described in section 3.1, can bridge a critical translational gap between \textit{in vitro} and animal models and humans by addressing discrepancies in pharmacokinetics. Second, multi-omics profiling provides an unbiased and comprehensive phenotype readout that contains information about target transferability\cite{khodosevich2024drug}, drug mode of action\cite{deepce_pham2021deep}, and pharmacokinetics\cite{chen2022integration}. Finally, techniques such as foundation models, contrastive learning, transfer learning, and other approaches that empower generative AI have shown promising results in predicting clinical drug responses based on disease models\cite{he2022cross, transpro_wu2023hierarchical, codeae_he2022context}. The development of the PVH will greatly benefit from substantial investments in improved disease models, high-throughput omics profiling, and data-efficient AI techniques. The Illustration is shown in Figure \ref{fig:main}

\section{Road map of developing PVHs}

\subsection{Challenges in developing the PHV}
While the rapidly increasing amount of omics data and improved disease models will facilitate developing the PVH, the gap between disease models and humans will always persist, as real-world human data for new molecular entities will be scarce or even completely unavailable. Consequently, new unseen cases (e.g., a chemical, a understudied protein, or a patient) may differ significantly from the data used to train the model, leading to an out-of-distribution (OOD) scenario\cite{liu2021towards}. Unfortunately, the OOD space for chemicals, biomolecules, and phenotypes in drug discovery is staggeringly vast\cite{wu2024ai,sasse2024unlocking,baek2024towards}, presenting a significant challenge in developing the PVH that is generalizable, trustworthy, and interpretable. 

A fundamental limitation of conventional machine learning techniques is their inability to handle small data or OOD cases. Although foundation models for biological entities\cite{li2024progress} may alleviate the small data and OOD problem, it persists when data is highly biased, in a zero-shot setting or answering "what-if" questions.  Additionally, in real-world applications, it is crucial to quantify the prediction uncertainty of new cases\cite{gawlikowski2023survey,seoni2023application}. This is particularly important in high-stakes areas like drug discovery and precision medicine. Existing statistics framework for uncertainty quantification typically assumes that data follow an identically and independently distributed (IID) pattern\cite{he2023survey}, but this assumption breaks down in OOD scenarios. Furthermore, model interpretability is vital for the PVH. Current state-of-the-art interpretation methods rely on mapping input features to output labels\cite{linardatos2020explainable}. However, in OOD cases, the "label" might be unobserved (e.g., a new cancer cell type), making existing methods potentially unreliable for explaining OOD predictions.


Here, we propose three complementary approaches to building generalizable, trustworthy, and interpretable PVH. This requires not only advancing machine learning techniques but also synergistically integrating AI, knowledge engineering, biophysics, and systems biology.

\subsection{Developing new machine learning techniques to address OOD challenges}

Causal representation learning may provide a foundation for building models that perform reliably even when faced with OOD data\cite{sanchez2022causal, michoel2023causal}. Traditional machine learning models typically learn spurious correlations that only hold in the training distribution. In contrast, causal representation learning focuses on disentangling the true causal factors that remain invariant across different environments or distributions. By learning these causal factors, models become more robust and generalize better to new, unseen distributions. Causal representation learning is also essential to develop explainable machine leaning techniques that can be extrapolated to OOD cases. 

The uncertainty quantification of point prediction is critical for cost- and risk-sensitive applications like drug discovery. New uncertainty quantification methods for drug discovery are needed to 1) improve the metrics, sampling, and clustering of multi-modal embedding space of the PHV, 2) integrate out-of-distribution detection and the statistics frameworks of uncertainty quantification such as conformal prediction\cite{angelopoulos2021gentle}, and 3) distinguish aleatoric uncertainty (also called statistical uncertainty mainly due to the noise of data) and epistemic uncertainty (also called systematic uncertainty from imperfect models). Recent studies in the uncertainty quantification of OOD binding affinity predictions using multi-modal models represent the effort along this direction\cite{Badkul2024.01.05.574359}.

\subsection{Integrating data-driven modeling and mechanism-based modeling}

Mechanism-based mathematical modeling emerges as a promising technique with the potential to address the challenges mentioned in machine learning. For instance, constraint-based metabolic network modeling has the capability to directly predict organismal phenotypes, such as growth rates under diverse conditions\cite{cortese2024applications}. Unlike "black box" machine learning models, the mechanism-based model explicitly represents the processes and interactions within a system, providing insights into the underlying principles governing the system. Consequently, the mechanism-based model can make predictions by leveraging existing knowledge, even in situations where data is scarce or difficult to acquire. Such a model exhibits greater generalizability, performing well across different scenarios. The transparency of the mechanism-based model also facilitates the interpretation of predictions and understanding the factors influencing outcomes, a crucial advantage for critical applications of the PVH. Furthermore, the mechanism-based model seamlessly integrates prior knowledge into the modeling process, enhancing the accuracy and relevance of predictions.

While mechanism-based mathematical modeling has its strengths, it comes with disadvantages and limitations when compared to machine learning. The mechanism-based model is typically designed based on existing knowledge and assumptions, and it may fail to capture and represent new or unexpected data patterns that extend beyond the assumed mechanisms. Additionally, a sophisticated model is often required to adequately represent the complexity of a biological system for making reliable predictions. However, solving complex mathematical equations derived from the mechanism-based model can be computationally demanding, thereby hindering the model’s practical application in a whole-person model of the PVH. Model parameterization poses another challenge for complex models, as parameters often need to be estimated from data, introducing uncertainties and potential inaccuracies. To make the mechanism-based model mathematically tractable, it often relies on simplifications and approximations of real-world systems, which can lead to inaccuracies, especially when dealing with highly complex or nonlinear systems.

Recently, efforts have been made to integrate mathematical modeling with machine learning for predicting cellular phenotypes. On one hand, leveraging biologically-informed data structures, such as genome-scale metabolic models\cite{vijayakumar2021protocol}, holds significant potential to both enhance the predictive power of AI-driven phenotype prediction models and provide insights into causal mechanisms at the meso-phenotype scale. On the other hand, novel deep learning techniques such as diffusion models and PINN show promise in addressing the computational complexity of mathematical modeling. For example, Lewis \& Kemp improved the accuracy of identifying patient subgroups by integrating genome-scale metabolic models with machine learning classifiers, offering a novel approach for personalized predictions of radiation response\cite{lewis2021integration}.

\subsection{Integrating molecular models with physiological and real-world models}


Data-driven or mechanism-based models built from real-world data (such as clinical data from EHRs, wearable device data, and lifestyle and behavioral information), are essential-but not sufficient-components for constructing PVHs for drug discovery and development. Integrating micro-scale molecular models with macro-scale human models is crucial for achieving this goal. 

Several promising directions have emerged for the integration of micro-scale models with macro-scale models. In purely data-driven models, cross-level end-to-end learning has been proposed to predict perturbed organismal phenotypes from genotypes\cite{wu2024ai}. For example, population clinical data can be embeded using unsupervised learning to enhance genomics discovery and prediction\cite{yun2024unsupervised}. When the macro-scale model is mechanism-based, outputs from micro-scale models can be used as inputs of the macro-scale model. For instance, machine learning can be used to predict the parameters for PBPK models of compounds during the early stage of drug discovery\cite{wu2024predicting}. Knowledge-based model integration may unify diverse models across scales, as prompted by NIH whole-person project\cite{nccih_whole_person_research}. 

\section{Conclusion}
The pharmaceutical industry faces unprecedented challenges in developing effective treatments for complex diseases, with conventional drug discovery approaches showing diminishing returns despite technological advances. The proposed PVH represents a paradigm shift in drug discovery and development, offering a potential solution to bridge critical translational gaps between molecular design and clinical efficacy. By enabling direct optimization of therapeutic outcomes in virtual human systems, PVH could fundamentally transform how we discover and develop drugs, particularly for complex, multi-factorial diseases.

The convergence of several technological advances in AI, omics techniques, and microphysiological systems makes this vision increasingly feasible. However, significant challenges remain in developing explainable, trustworthy and generalizable PVH. The vast out-of-distribution space in drug discovery, the inherent complexity of human physiology, and the persistent gap between disease models and human systems require novel solutions. We propose a three-pronged approach to address these challenges: (1) developing new machine learning techniques specifically designed to handle out-of-distribution predictions, uncertainty quantification, and intepretability, (2) integrating data-driven approaches with mechanism-based modeling to enhance generalizability and interpretability, and (3) bridging the scales between molecular-level models and physiological/real-world models.
 
\section*{Acknowledgement}
This project has been funded with federal funds from the National Institute of General Medical Sciences of the National Institute of Health (R01GM122845), the National Institute on Aging of the National Institute of Health (R01AG057555, R21AG083302), and the National Science Foundation (2226183).

\newpage
\bibliographystyle{ieeetr}
\bibliography{ref}

\end{document}